# Numerical Investigation of Natural Convection Dynamics in Open-Ended Vertical Channels with Different Aspect Ratios

Heba A. Alaaeldin[1], Ahmed A. Hamada[2], Mohamed A. El-Naggar[1,3], Elsayed Z. El-Ashtoukhy[1], and Mahmoud M. Taha[1]

[1]Chemical Engineering Department, Faculty of Engineering, Alexandria University, Alexandria 21544, Egypt.
[2]Mechanical Engineering Department, The University of Texas at Dallas, Richardson, TX 75080, USA.
[3]Department of General Subjects, University of Business and Technology, Jeddah 21432, Saudi Arabia.

Corresponding author: Heba A. Alaaeldin (heba.alaa@alexu.edu.eg)

*Abstract*—Understanding the thermal and flow fields associated with natural convection is of great importance in the design and manufacture of passive cooling systems, solar collectors, and ventilation systems. The aim of this study is to numerically investigate the naturally driven flow behaviour in open-ended vertical channels. Specifically, laminar airflow (Rayleigh number of $5 \times 10^5$) is naturally driven through a vertical channel that is asymmetrically heated with a uniform heat flux of ($2043 \, w/m^2$). The compressible OpenFOAM solver (buoyantPimpleFoam) is employed, and open boundary conditions are applied at both the inlet and outlet to account for the effect of ventilation. The results exhibit good agreement with a numerical benchmark. Furthermore, a study of the airflow through two vertical channels, with two different aspect ratios of 5 and 31.25, is conducted. The results indicate that, in both cases of aspect ratios, the flow accelerates near the heated wall and decelerates near the adiabatic wall, forming a near-constant velocity region around the channel centreline. Increasing the aspect ratio diminishes the reverse flow. However, the flow pattern near the heated wall remains largely unaffected by aspect ratio changes. Additionally, the fluid behaviour downstream significantly affects velocity measurements at the inlet.

*Keywords—laminar flow, natural convection, vertical channel, open boundary conditions, OpenFOAM.*

I. INTRODUCTION

Natural convection is a heat transfer mechanism that occurs due to buoyancy forces, which are stimulated as a result of thermally induced density variations in a gravitational field. When a fluid is heated, it expands and becomes less dense; thus, buoyancy forces overcome fluid inertia and frictional resistance, causing the hot fluid to rise, while the cooler fluid sinks [1]. The resulting motion of the fluid is called natural convection flow. Indeed, this motion persists as long as the temperature gradient is maintained [2].

In general, natural convection is characterised by its significance in energy saving, utilization of environmentally friendly materials, and low-cost components. Therefore, studying natural convection in vertical channels is interesting from an applied perspective as it is used in many practical heat transfer applications, such as the flow between double-glazed windows [3, 4], ventilation systems [5-8], solar collectors [9-11], solar dryers [12], natural-convection air cooling of electronic equipment [13-18], and its role as a passive safety feature in advanced very high-temperature reactors [19-21].

Elenbaas [22] carried out a detailed study of the thermal characteristics of cooling by natural convection in smooth, parallel-walled vertical channels. This pioneering experimental work laid the foundation for subsequent experimental and numerical studies of natural convection in vertical channels under different heating conditions using different measurement techniques. For instance, Ji et al. [23] numerically studied the impact of domain extension on natural convection heat transfer in a vertical rectangular channel. They utilized the realizable k-ε turbulence model for their simulations. Their findings revealed that increasing the extended domain resulted in an approximately 8% rise in mass flow rate and up to a 13.5% increase in the heat transfer coefficient. Their work serves as a reference for selecting appropriate extended domains in natural convection scenarios, and they also identified the optimal parameters for the extended domain. Pallares et al. [24] conducted direct numerical simulations (DNS) of fully developed turbulent natural convection airflow at relatively low Rayleigh numbers in an asymmetrically heated vertical channel. They found that under fully developed flow conditions, the shear stress is higher on the heated wall compared to the adiabatic wall. However, near the insulated wall, turbulent intensities and shear stress exhibit greater values. In contrast, temperature fluctuations and turbulent heat fluxes near the heated wall exhibit greater intensity than those near the adiabatic wall. Additionally, turbulent shear stress plays a central role in generating turbulent kinetic energy inside the asymmetrically heated channel, similar to forced convection channel flow. Zoubir et al. [25] investigated the applicability of using water to predict natural convection in an air-filled channel, particularly in the context of thermal stratification. They studied laminar natural convection within a vertically oriented channel inside an enclosure, where the channel walls were symmetrically heated with uniform heat flux. By solving the 2D unsteady Navier-Stokes equations using the Finite Difference Method, they compared the flow and heat transfer characteristics of water and air, considering the influence of external stratification. Their research revealed that the effect of thermal stratification on heat transfer by natural convection is greater when air is employed as the working fluid. In addition, they provided some explanations for the disparities observed in the literature between numerical and experimental results when examining water and air as working fluids. Njoroge and Gao [26] conducted numerical investigations to examine how the shape of adiabatic extensions influences heat transfer. Specifically, they focused on wall temperatures, mass flow rates, and Nusselt numbers. Their findings contribute to advancements in the design of air-cooled passive systems, particularly in the context of modern nuclear reactors. Kim et al. [27] conducted an experimental study to examine turbulent natural convection between open-ended vertical parallel plates subjected to asymmetric heating conditions. They analysed the ratio of convective to radiative heat transfer rates on the wall that was directly heated. Based on their experimental findings, they developed two correlations to describe the heat transfer due to natural convection between the vertical parallel plates. Cherif et al. [28] carried out an experimental study to investigate the thermal and dynamic characteristics of airflow enhanced by natural convection in an open vertical channel. The particle image velocimetry (PIV) method was employed to visualize the flow. Their examination focused on how the Rayleigh number and the aspect ratio affect velocity profiles, as well as convective and radiative heat transfer. Desrayaud et al. [29] examined four variant sets of open boundary conditions (BCs) used in the numerical analysis of laminar flow enhanced by natural convection in a vertical asymmetrically heated channel. The channel had no upstream or downstream expansions. Results highlighted that the local and average Nusselt numbers based on the entry temperature are mainly unaffected by the type of BCs applied at the inlet section. Additionally, the net fluid flow rate through the channel and the characteristics of the recirculation cells exhibited variations with the examined BCs. However, despite these alterations in flow patterns, there was no significant impact on the fluid flow rates exiting the channel through the exit section. This study provided a benchmark solution for validating the numerical schemes in open-ended vertical channels.

The present study aims to investigate the laminar natural convection flow fields in vertical open-ended asymmetrically heated channels. Air is employed as the working fluid ($Ra = 5 \times 10^5$) inside an open channel that is one-sided heated with a uniform heat flux of $q = 2043\ w/m^2$. Firstly, the results obtained using the buoyantPimpleFoam solver are compared with the benchmark work [29]. Then, the results of investigations of the effect of two different channel's aspect ratios on velocity fields and reversed flow is discussed. The outcomes from this work are valuable for analysing and designing natural convection ventilation systems using OpenFOAM in future research.

## II. PHYSICAL MODEL

### A. Problem Definition

In this study, the vertical channel is represented by two parallel vertical walls separated by a gap width of 0.016 m. The study investigates two different channels with varying heights, one channel measures 0.16 m, while the other stands at 1 m. As shown in Fig. 1, a uniform heat flux ($q$) of $2043\ w/m^2$ is applied to the middle half section of the left-side wall, while the bottom and top quarters of this channel, as well as the right-side wall, remain adiabatic. Air ($Pr = 0.71$) naturally flows upward through the channel, directed toward the top exit, following its heating by the applied heat flux. In the present model, it is assumed that the fluid flow is two-dimensional and laminar ($Ra = 5 \times 10^5$). Additionally, the effects of thermal radiation and heat conduction within the internal solid walls are neglected.

### B. Governing Equations

The governing equations of fluid dynamics and heat transfer, namely mass conservation, momentum conservation, and energy conservation are [30]:

$$\frac{\partial \rho}{\partial t} + \nabla \cdot (\rho U) = 0, \qquad (1)$$

$$\frac{\partial \rho U}{\partial t} + \nabla \cdot (\rho U U) = -\nabla P_{rgh} - g \cdot r\, \nabla P + \nabla \cdot \tau, \qquad (2)$$

$$\frac{\partial \rho h}{\partial t} + \nabla \cdot (\rho U h) + \frac{\partial \rho K}{\partial t} + \nabla \cdot (\rho U K) - \frac{\partial P}{\partial t} = -\nabla \cdot q + \nabla \cdot (\tau \cdot U) + \rho r + \rho\, g \cdot U. \qquad (3)$$

where $\rho$ is mass density, $U$ is the velocity, $K$ is the kinetic energy per unit mass, ($K = |U|^2/2$), $g$ is the gravitational acceleration, $\tau$ is the viscous stress tensor, $q$ is the heat flux vector, defined as positive inwards, $h$ is the enthalpy, ($P_{rgh} = P - \rho\, g\, r$) where $P$ is the static pressure field and $r$ is the position vector.

## III. COMPUTATIONAL MODEL

### A. Numerical Implementation

buoyantPimpleFoam is an OpenFOAM® v2012 transient solver for modelling buoyant, laminar, and turbulent flows of compressible fluids in scenarios related to ventilation and heat transfer [31]. The discretisation scheme used is Euler for solving the unsteady term in the momentum and energy conservation equations. Moreover, the Gauss linear scheme is used for all gradient terms. For the divergent terms, the Gauss upwind scheme is selected. In addition, the Laplacian terms are handled with Gauss linear correction [32].

### B. Computational Domain and Boundary Conditions

The computational domain was discretised using a two-dimensional mesh of quadrilateral cells, as demonstrated in Fig. 2. For the entire domain, the elements' sizes are the same.

The thermal boundary condition is set as a fixed heat flux ($q = 2043\ w/m^2$) on the middle half of the left wall. The remaining walls are considered adiabatic *(zeroGradient)*. Hence, the velocity is unknown at the inlet of the channel; entrainment boundary conditions are suitable for simulating naturally driven airflow. In this approach, the velocity is allowed to find its value by combining the *(prghTotalPressure)* condition of the pressure field, where the hydrostatic component is subtracted based on a height ($\Delta h$) above some reference level, $P_{rgh} = P - \rho|g|\Delta h$ with the velocity condition *(PressureInletOutletVelocity)*. The relationship between velocity and pressure follows the generalized Bernoulli theorem:

$$P_\circ + 0 = P_\circ - \rho|g|\Delta h - \frac{1}{2}\rho|u^2| \qquad (4)$$

The left-hand side of the equation represents the conditions of the fluid flow before it enters the channel, where $P_\circ$ is the static pressure (1 bar) and (zero) refers to the assumed value of the initial velocity, whereas the right-hand side of the equation moderates the rise in the dynamic pressure ($\frac{1}{2}\rho|u^2|$) by the decrease in the static pressure [33]. The inlet and outlet boundary conditions for both temperature and pressure (P) are specified as *(inletOutlet)*. Outlet conditions are set as *(zeroGradient)* for velocity and *(inletOutlet)* for $P_{rgh}$. At the solid walls, no-slip conditions are imposed on the velocity field, and the $P_{rgh}$ is specified as *(fixedFluxPressure)* to consider the gravitational body force and no-slip conditions of the velocity field.

*C. Grid Convergence*

A mesh-independent study was carried out for the high aspect ratio (31.25) channel by decreasing $\Delta x$ and $\Delta y$ of cells while maintaining the same aspect ratio, as illustrated in Fig. 3. Different mesh sizes were employed, as demonstrated in Tab. I. The maximum value of the Courant-Friedrichs-Lewy (CFL) number was set to 1.5, and the time step size varied accordingly in each simulation. The results shown in Fig. 3 reveal that the grid ($40 \times 2500$) is precise enough to produce sufficiently accurate results. Consequently, this grid was used for all computations.

IV. VALIDATION

The selected OpenFOAM solver is validated through an analysis of momentum, pressure, and thermal results, using the benchmark solution provided in Ref. [29]. The local Bernoulli relation (LB) and the global Bernoulli relation (GB) represent two types of pressure boundary conditions arising from Bernoulli's theorems. These conditions are defined and investigated in the benchmark study [29]. The validation process utilizes the same geometry details and boundary conditions as the reference study. Additionally, a uniform grid consisting of ($100 \times 1000$) cells is employed.

Fig. 4 shows the comparison of the variation of the non-dimensional vertical velocity component ($w$) across the horizontal axis at different heights of the channel in the current study with the reference solution. The computational method shows good agreement with the benchmark results.

Both $w$ and dimensionless pressure variations along the vertical axis at the mid-width are also plotted, as shown in Figs. 5 and 6, respectively. The dimensionless pressure is calculated using the equation defined in the benchmark problem, $P(x, A) = -\frac{1}{2}[w(x, A)]^2$. The current results exhibit consistency with those of the benchmark study [29] up to 75% of the channel's height. However, distinct trends are observed in the top 25% section.

Furthermore, the heat transfer validation results, as listed in Tabs. II-IV, were obtained by comparing three dimensionless parameters $\theta_b(z)$, $\widetilde{Nu}_1(z)$, and $\widetilde{Nu}_2(z)$, where $\theta_b(z)$ represents the dimensionless bulk temperature parameter, $\widetilde{Nu}_1(z)$ is the inverse of the temperature at the left wall, and $\widetilde{Nu}_2(z)$ is the inverse of the difference between the temperature at the left wall and the bulk temperature, as defined in the benchmark study [29]. Overall, the current results show good agreement with all examined parameters, except for the calculations of $\widetilde{Nu}_2$ in the top quarter of the channel. This discrepancy was also observed in the benchmark study.

V. RESULTS AND DISCUSSION

Velocity measurements were conducted for laminar natural convection flow with a Rayleigh number (Ra) of $5 \times 10^5$ and a Prandtl number (Pr) of 0.71 in two asymmetrically heated narrow vertical channels. Both channels are identical but have different aspect ratios: one with a low aspect ratio of 5 and the other with a high aspect ratio of 31.25. The aspect ratio is calculated for the middle half section of the channel, where a heat flux of $q = 2043\ w/m^2$ is applied, while the first and last quarters of the channel remain adiabatic to prevent airflow instability in those regions. The initial velocity of the inlet airflow is ($0.01\ m/s$) in both cases. The results in this section are obtained using mesh ($100 \times 1000$) for the low aspect ratio case and mesh ($40 \times 2500$) for the high aspect ratio case.

Fig. 7 shows the vertical velocity profiles within the low aspect ratio channel. The observed flow pattern indicates an upward motion across the channel, maintaining a consistent velocity at the entry adiabatic region, with high-velocity gradients along the shear layer adjacent to the channel walls. At a height of 0.04 m, the velocity distribution exhibits symmetric behaviour around the channel centreline. While the velocity reaches its maximum value at the centre, it gradually decreases as it approaches the channel walls due to viscous effects. At a height of 0.08 m, the velocity profile exhibits an increase in velocity toward the heated wall, which is attributed to the higher flow temperature and the resulting buoyancy force within the heated middle section. On the other hand, velocity decreases near the adiabatic wall. The flow patterns across the channel at heights of 0.12 m and 0.16 m exhibit similar trends. The flow accelerates upward near the hot side and decelerates as it moves downward near the adiabatic side, indicating the presence of a reverse flow in the upper section of the channel. Because of the momentum energy exchange between the reverse flow and the ascending heated airflow, the zero-velocity point of the reverse airflow moves closer to the adiabatic plate as it goes down. The current results agree with the literature [34, 35].

The vertical velocity profiles for the high aspect ratio case are shown in Fig. 8. The velocity distribution at the inlet reveals that the flow ascends through the channel at a nearly constant value, whereas it has a high velocity gradient near the walls due to the viscous effect. The velocity distribution across the channel exhibits a parabolic shape at a height of 0.25 m. Notably, closer to the channel walls, the velocity gradients decrease due to the growing thickness of the boundary layer as the flow moves upward. When the flow reaches a height of 0.5 m, a significant velocity gradient develops on the shear side near the hot wall. This is attributed to the rise in flow temperature, resulting in the development of buoyancy forces within the left middle region. Meanwhile, the velocity experiences a slight decrease in the other half of the channel near the adiabatic side. The velocity distributions across the channel at heights of 0.75 m and 1 m exhibit similar patterns. The flow velocity greatly increases near the hot plate, while it undergoes a slight deceleration near the adiabatic plate.

Upon comparing the velocity profiles depicted in Figs. 7 and 8, it is revealed that, overall, the airflow velocities in the high aspect ratio case are higher than those in the low aspect ratio case due to the greater heat quantity gained near the hot plate, and therefore the enhanced buoyance force is greater. However, the flow pattern near the hot plate is not significantly influenced by increasing the aspect ratio. In both cases, the maximum velocity shifts away from the hot wall, indicating the gradual thickening of the boundary layer as the flow progresses downstream. In both scenarios, the flow accelerates near the hot side of the channel, creating a near-constant velocity region around the centreline. Specifically, in the case of a high aspect ratio channel, the near-constant velocity region is observed near the adiabatic wall, whereas in the low aspect ratio channel, it exists near the hot wall due to the presence of reverse flow in the upper section of the channel. The visualization of this behaviour is depicted by the velocity streamlines shown in Figs. 9 and 10 for both cases. Although the airflow has the same Rayleigh number in both cases, changing the aspect ratio significantly impacts the flow behaviour in the upper section of the channel. Specifically, as the aspect ratio increases, the reverse flow diminishes. The current results coincide with the previous work [27]. Additionally, the velocity profiles within the first adiabatic quarter of the channel exhibit similar trends for both cases. However, the velocity values in the high aspect ratio scenario are approximately six times greater than those in the low aspect ratio case. Notably, despite identical inlet conditions, changes in the downstream significantly influence the inlet flow.

## VI. Conclusion

The current study aims to perform numerical investigations on the flow behaviour driven by natural convection in open-ended vertical channels. The results demonstrate good agreement with the benchmark study mentioned earlier. Additionally, an examination of the fluid behaviour of laminar natural convection airflow ($Ra = 5 \times 10^5$) in two vertically asymmetrically heated channels with two different aspect ratios of 5 and 31.25 is performed. Our findings reveal that increasing the aspect ratio significantly impacts the flow pattern, leading to a diminishment of the reverse flow. Furthermore, velocity measurements at the inlet section are markedly influenced by the fluid behaviour in the downstream direction.

TABLE I. CHARACTERISTICS OF MESH INDEPENDENT TEST

| Mesh size | $\Delta x$ (Or) $\Delta y$ | Simulation Time | $v$ | Error % |
|---|---|---|---|---|
| 10 × 625 | 0.0016 m | 0.33 hrs | 1.11 | 5.34 |
| 20 × 1250 | 0.0008 m | 2.5 hrs | 1.08 | 2.32 |
| 30 × 1875 | 0.00053 m | 8.8 hrs | 1.07 | 1.07 |
| 40 × 2500 | 0.0004 m | 28 hrs | 1.05 | --- |

TABLE II. VALUES OF $\theta_b(z)$, THE DIMENSIONLESS BULK TEMPERATURE PARAMETER

| $\theta_b(z)$ | Benchmark solution | Current study | Deviation % |
|---|---|---|---|
| $\theta_b(z = 3A/8)$ | 0.01463 - 0.02094 | 0.02222 | 6.11 |
| $\theta_b(z = A/2)$ | 0.02919 - 0.04182 | 0.04655 | 11.31 |
| $\theta_b(z = 5A/8)$ | 0.04375 - 0.06270 | 0.07212 | 15.02 |
| $\theta_b(z = 3A/4)$ | 0.05820 - 0.08338 | 0.09773 | 17.21 |
| $\theta_b(z = 7A/8)$ | 0.05828 - 0.08356 | 0.09038 | 8.16 |
| $\theta_b(z = A)$ | 0.05823 - 0.08348 | 0.08705 | 4.28 |

TABLE III. VALIDATION OF $\widetilde{Nu}_1(z)$ CALCULATIONS

| $\widetilde{Nu}_1(z)$, $Nu_1(z \leq 3A/4)$ | Benchmark solution | Current study | Deviation % |
|---|---|---|---|
| $\widetilde{Nu}_1(z = 3A/8)$ | 7.198 - 7.325 | 6.406 | 11.00 |
| $\widetilde{Nu}_1(z = A/2)$ | 6.170 - 6.265 | 5.636 | 8.65 |
| $\widetilde{Nu}_1(z = 5A/8)$ | 5.627 - 5.716 | 4.825 | 14.25 |
| $\widetilde{Nu}_1(z = 3A/4)$ | 5.602 - 5.694 | 4.864 | 13.17 |

TABLE IV. CALCULATIONS OF $\widetilde{Nu}_2(z)$

| $\widetilde{Nu}_2(z)$, $Nu_2(z \leq 3A/4)$ | Benchmark solution | Current study | Deviation % |
|---|---|---|---|
| $\widetilde{Nu}_2(z = 3A/8)$ | 8.207 - 8.469 | 9.033 | 6.66 |
| $\widetilde{Nu}_2(z = A/2)$ | 7.671 - 8.311 | 8.187 | 1.49 |
| $\widetilde{Nu}_2(z = 5A/8)$ | 7.626 - 8.690 | 7.752 | 10.79 |
| $\widetilde{Nu}_2(z = 3A/4)$ | 8.528 - 10.52 | 10.006 | 4.89 |

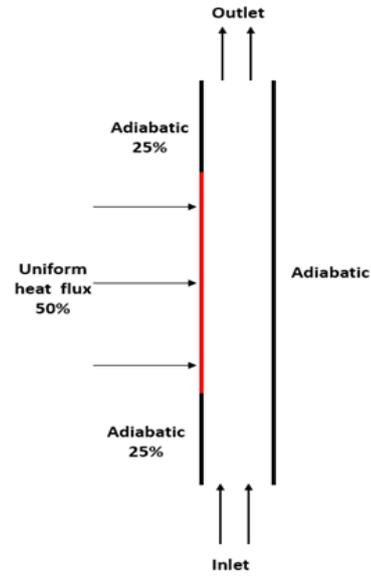

Fig. 1. Case description.

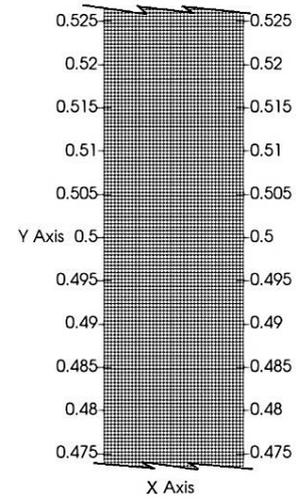

Fig. 2. The middle section of the channel demonstrates the uniform 2-D grid type used.

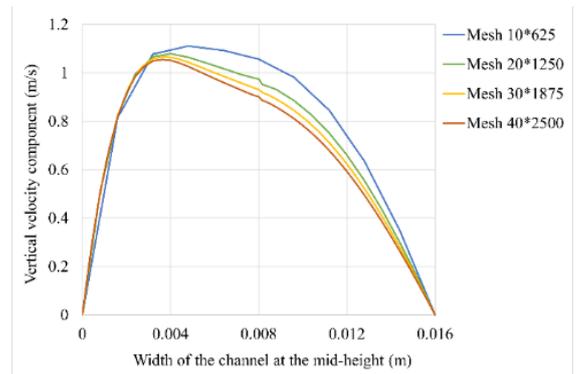

Fig. 3. Effect of different mesh sizes on radial distribution of vertical velocity component.

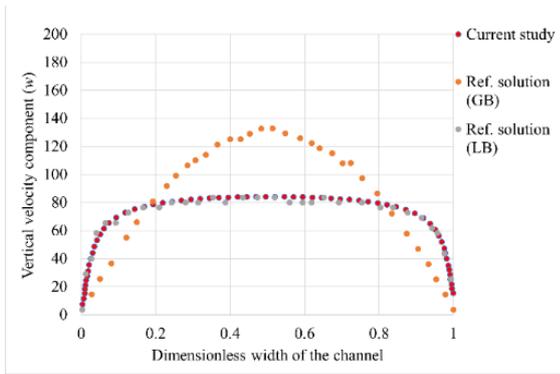

(a) Channel's dimensionless height = 0

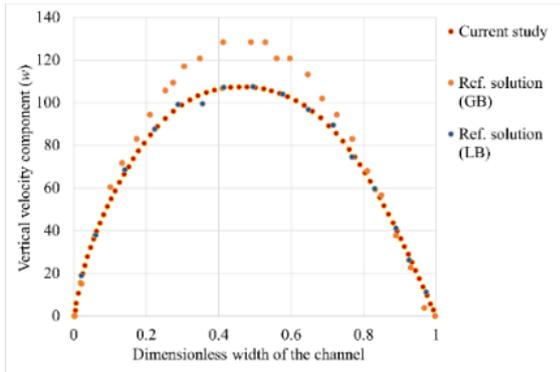

(b) Channel's dimensionless height = 0.25

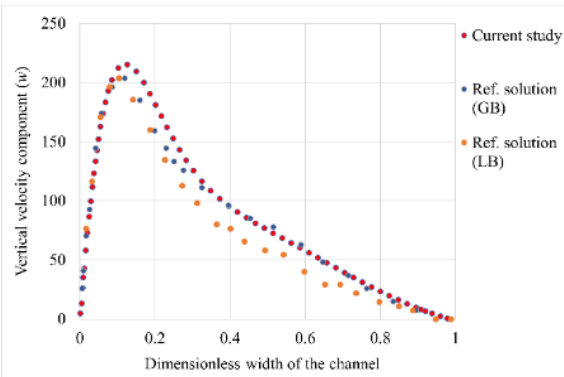

(c) Channel's dimensionless height = 0.5

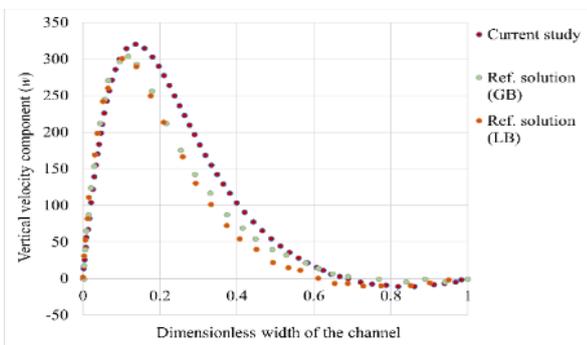

(d) Channel's dimensionless height = 0.75

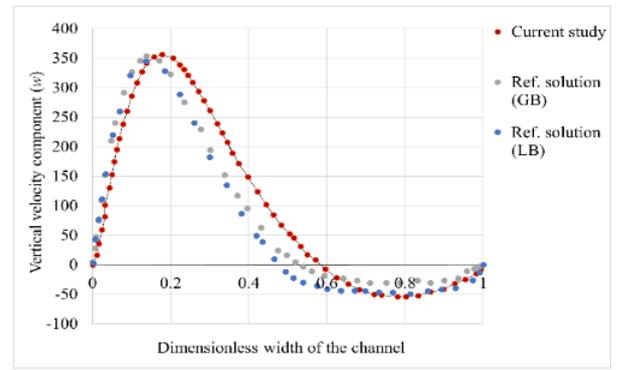

(e) Channel's dimensionless height = 1

Fig. 4. Comparison between the results conducted by the scheme utilized in the current study against benchmark solution [29].

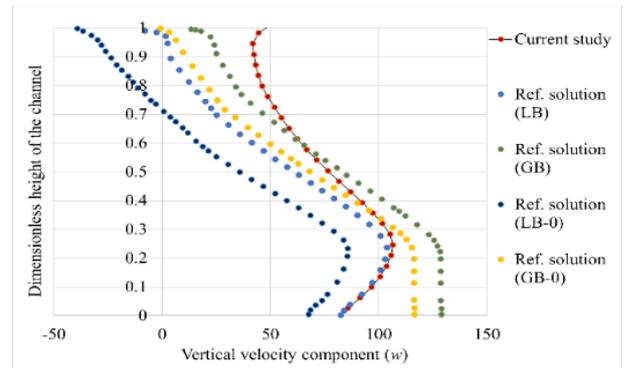

Fig. 5. Distribution of the non-dimensional vertical velocity component.

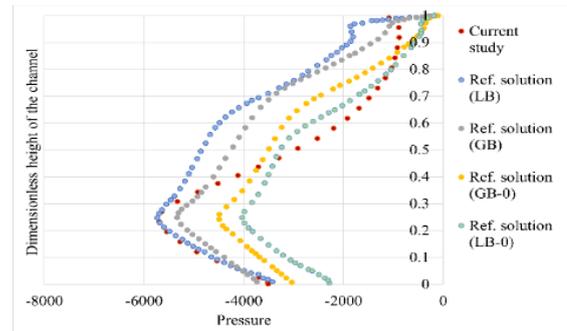

Fig. 6. Verification of current results against Desrayaud et al.[29].

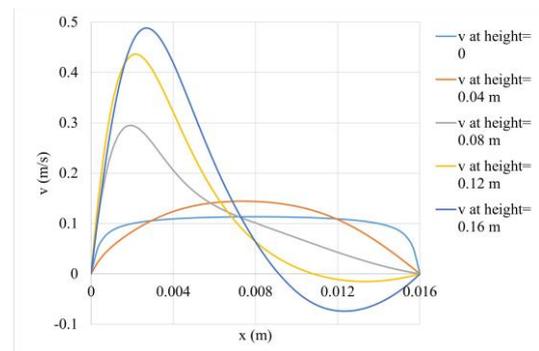

Fig. 7. Vertical velocity variation over the horizontal axis (x) at different heights for the case of the low aspect ratio.

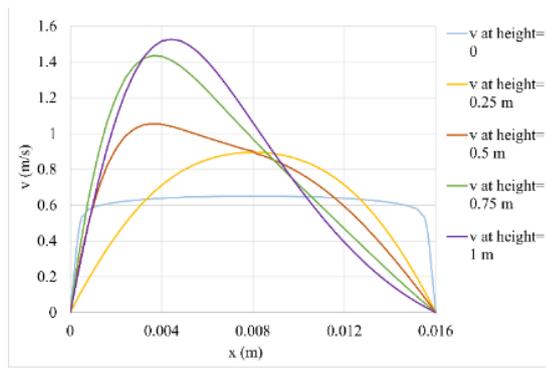

Fig. 8. Vertical velocity variation over the horizontal axis (x) at different heights for the case of the high aspect ratio.

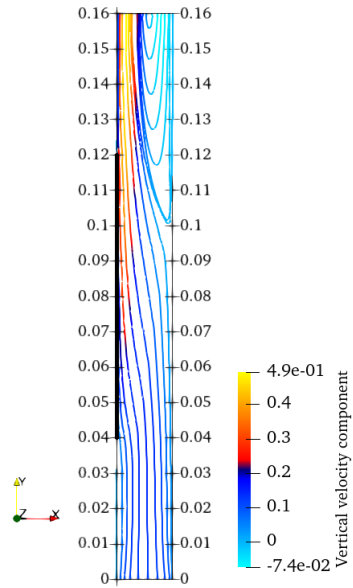

Fig. 9. Streamlines of the velocity in the low aspect ratio channel.

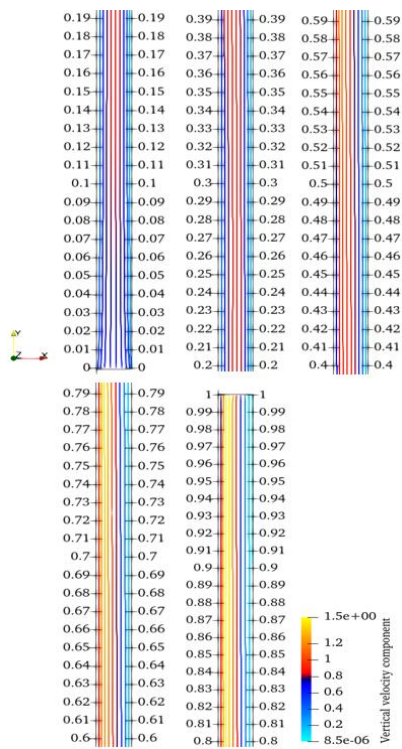

Fig. 10. Streamlines of the velocity in different sections of the high aspect ratio channel.